\documentclass[prd,preprintnumbers,superscriptaddress,amsmath,amssymb,nofootinbib,twocolumn,floatfix]{revtex4-2}

\usepackage{setspace}
\usepackage{graphicx}
\usepackage{hyperref}
\usepackage{booktabs}    % For improved table rules
\usepackage{units}
\usepackage{nameref}
\usepackage{amsmath,amssymb, mathtools}

\usepackage{xcolor}

\newcommand{\be}{\begin{equation}}
\newcommand{\ee}{\end{equation}}

\begin{document}

\preprint{KCL-PH-TH/2025-34}

\title{GW231123: Binary Black Hole Merger or Cosmic String?}% Force line breaks with \\

\author{Iuliu Cuceu}
 \email{icuceu@oca.eu}
 \affiliation{
 Université Côte d’Azur, Observatoire de la Côte d’Azur, CNRS, Laboratoire Artemis, 06300 Nice, France}
\author{Marie Anne Bizouard}%
 \email{marieanne.bizouard@oca.eu}
  \affiliation{
 Université Côte d’Azur, Observatoire de la Côte d’Azur, CNRS, Laboratoire Artemis, 06300 Nice, France}
\author{Nelson Christensen}
 \email{nelson.christensen@oca.eu}
 \affiliation{
 Université Côte d’Azur, Observatoire de la Côte d’Azur, CNRS, Laboratoire Artemis, 06300 Nice, France}
\author{Mairi Sakellariadou}
\email{mairi.sakellariadou@kcl.ac.uk}
\affiliation{Theoretical Particle Physics and Cosmology Group, Physics Department, King’s College London,
University of London, Strand, London WC2R 2LS, United Kingdom\\ }
\affiliation{
 Université Côte d’Azur, Observatoire de la Côte d’Azur, CNRS, Laboratoire Artemis, 06300 Nice, France}

\date{\today}% It is always \today, today,

\begin{abstract}
The LIGO-Virgo-KAGRA Collaboration recently reported an exceptional gravitational-wave event, GW231123. This gravitational-wave signal was assumed to be generated from the merger of a binary black hole system, with source frame masses of $137^{+22}_{-17}~\textup{M}_\odot$ and $103^{+20}_{-52}~ \textup{M}_\odot$ (90\% credible intervals). As seen by the two LIGO detectors, the signal has only $\sim 5$ cycles, between 30 and 80 Hz, over $\sim 10$ ms. It is of critical importance to confirm the origin of this signal. Here we present the results of a Bayesian model comparison to test whether the gravitational-wave signal was actually generated by a binary black hole merger, or emitted from cusps or kinks on a cosmic string. We find significant evidence for a binary black hole merger origin of the signal.

\end{abstract}

\maketitle

%\tableofcontents

\section{\label{sec:level1}Introduction}
%The gravitational-wave detectors Advanced LIGO~\cite{LIGOScientific:2014pky}, Advanced Virgo~\cite{VIRGO:2014yos} and KAGRA~\cite{KAGRA:2021duu} (LVK) are currently in their fourth period of observation, O4, which commenced in May 2023. 
The LIGO~\cite{LIGOScientific:2014pky}, Virgo~\cite{VIRGO:2014yos}, KAGRA~\cite{KAGRA:2020tym} Collaboration (LVK) has recently announced the detection of a intriguing gravitational-wave event, GW231123~\cite{LIGOScientific:2025rsn}.
GW231123 was observed by the two LIGO detectors at 13:54:30.634 UTC on November 23, 2023. 
Assuming that the signal comes from a compact binary coalescence, the LVK finds
that this gravitational-wave event was produced by a binary black hole merger, with source frame masses of $137^{+22}_{-17}~ \textup{M}_\odot$ and $103^{+20}_{-52}~ \textup{M}_\odot$ (90\% credible intervals). Such a massive binary black hole system poses many questions as to its formation. However, to determine that GW231123 is produced from a binary black hole is made difficult by the fact that only $\sim$ 5 wave cycles are observed. This signal is present between 30 and 80 Hz, lasting $\sim 10$ ms in the LIGO data~\cite{LIGOScientific:2025rsn}. 

It is probable that in the future there will be other gravitational-wave signals of short duration. More massive binary black hole systems will be observable over shorter times and smaller frequency bands. It will be important to have methods for quantifying the probabilities of different source models. We do this here, using different cosmic string emission 
channels, plus a more generic power-law signal emission. Ultimately other short duration transient models will need to be included in model comparison studies.
While the LVK makes the assumption that GW231123 comes from a binary black hole, 
it acknowledges other possible interpretations for the signal; 
eccentricity, gravitational lensing, core-collapse supernova, exotic compact objects, or cosmic strings. Since these alternative scenarios are not tested in~\cite{LIGOScientific:2025rsn},
we investigate here whether a cosmic string could be the source of the GW231123 event.

Cosmic strings are 1+1 dimensional topological defects, generically predicted in beyond the Standard Model particle physics~\cite{Jeannerot:2003qv}.
Gravitational waves emitted by cosmic strings provide one of their most promising observational signatures  accessible by ground-based interferometers~\cite{LIGOScientific:2017ikf,LIGOScientific:2021nrg}. Cosmic strings can be either super-horizon (long) or sub-horizon (loops). At high frequencies, the gravitational-wave spectrum of an oscillating loop is dominated by bursts emitted by cusps and kinks~\cite{Damour:2000wa,Damour:2001bk,Damour:2004kw}.
Cusps are short-lived features on a string loop that briefly travel at the speed of light. The number of cusps per loop has not yet been determined; they are generic features for smooth loops.
Kinks are discontinuities in the tangent vector of the string that propagate
at the speed of light. 
Kinks, appearing in pairs, are the result of string intercommutations and therefore exist on long strings as well. Numerical experiments of Nambu-Goto strings (the width is negligible as compared to its size) concluded that kinks accumulate over the cosmological evolution~\cite{PhysRevLett.60.257,PhysRevLett.64.119,PhysRevD.42.349}.
Since long strings also have kinks, they can also emit gravitational waves ~\cite{Sakellariadou:1990ne}. Cusps create beamed gravitational waves in the forward direction of the cusp, while kinks produce gravitational
waves with fan-like emission. In contrast, kink-kink collisions emit gravitational waves isotropically.

Previously, LIGO and Virgo reported the detection of a gravitational-wave signal, GW190521, which they also stated came from a binary black hole system~\cite{LIGOScientific:2020iuh,LIGOScientific:2020ufj}. Model comparison study results were presented, showing
that the binary black hole merger scenario was favored over cosmic strings\footnote{Another study considered a particular case of a planar circular string loop collapsing to a black hole~\cite{Aurrekoetxea:2023vtp}.}.

In what follows, we present a Bayesian model comparison between a binary black hole and a cosmic string as potential sources of the GW231123 event.
We conclude that the binary black hole origin of GW231123 is strongly favored over cosmic string.

\section{Method}
The data from the two LIGO detectors for the GW231123 event, from which identified noise such as the 60 Hz power mains and its harmonics has been subtracted, is publicly available at the Gravitational Wave Open Science Center~\cite{gwoscGW231123,KAGRA:2023pio}.
To compare a binary black hole versus a cosmic string as the origin of the GW231123 event, we performed a Bayesian study~\cite{Christensen:2022bxb}.  
We thus use Bayes factors as a discriminant for the models considered.
\begin{equation}
    \mathcal{B}\mathcal{F}=
    \frac{\mathcal{Z}(d\vert {\cal M}_1)}{\mathcal{Z}(d\vert {\cal M}_2)}~,
\end{equation}
where the evidence $\mathcal{Z}$ is estimated with the nested sampling~\cite{2004AIPC..735..395S,10.1214/06-BA127} package \texttt{dynesty}~\cite{2020MNRAS.493.3132S,sergey_koposov_2023_7600689}, within the gravitational wave inference pipeline \texttt{bilby}~\cite{2019ApJS..241...27A,2020MNRAS.499.3295R}. ${\cal M}_1$  represents the model with a gravitational wave present in the detector noise, while ${\cal M}_2$ represents only the detector noise. We do not attempt to quantify the prior odds of a given model, so in effect, both a binary black hole merger and a cosmic string origin are equally likely a priori.

We consider a linearly polarized waveform expressed in the frequency-domain as
\be
h_i = {\cal A}_i  \Theta(f-f_{\rm low}) \Theta(f_{\rm high}-f) e^{-2\pi i f t_A} f^{-q_i} ~,
\ee
where $i= \{ \rm{c, k, kk, pl} \}$ denotes respectively the cusp, kink, kink-kink collision cases and a generic power-law to capture a range of possible sources. The power-law indices $q_i$ depend on the model, with~\cite{Damour:2000wa,Damour:2001bk,Damour:2004kw} $q_{\rm c}=4/3, q_{\rm k}= 5/3$, $q_{\rm kk}=2$ and $q_{\rm pl}$ kept free. The function ${\cal A}_i$ stands for amplitude, $t_A$ is the arrival time defined relative to trigger time (1384782888.634 GPS time) in Figs.~\ref{fig:cornerBBH}, \ref{fig:CScorner}
below,
$f$ denotes the frequency and $\Theta$ is the Heaviside function. The low frequency cutoff $f_{\rm low}$ is fixed at $20$ Hz in this analysis to take into account the large rise of the noise at low frequencies. The high frequency cutoff $f_{\rm high}$ is a free parameter that, in the case of the cusp and kink models, is inversely proportional to the cube of the beaming angle, which is defined as the angle between the line of sight and the emission cone axis. 

For the cosmic string models, we consider priors inspired from~\cite{Divakarla:2019zjj}.
Specifically, the ${\cal A}_i$ samples are drawn from a log-uniform distribution, from $10^{-23}\,\text{to}\,10^{-18}$ and the $f_{\rm high}$ is uniformly distributed starting from $25\,\text{Hz}$, ending at $448\,\text{Hz}$ for the cusp and $2000\,\text{Hz}$ for the other cosmic string cases. The temporal and sky localization priors match~\cite{LIGOScientific:2025rsn}. For the arbitrary power-law spectral index, the prior is uniformly distributed between three different choices for the lower bound, to determine the impact of the prior choice on the posterior distribution, and 4 as the upper bound.

For the the binary black hole waveform approximant, we take \texttt{NRSur7Dq4}~\cite{Varma:2019csw}, appropriate for the highly spinning high-mass compact binary coalescence scenario. We note that the LVK study used five different waveforms, but NRSur7Dq4 performed best in a signal injection study~\cite{LIGOScientific:2025rsn}. Although the choice of waveform approximant affects the parameter estimation scheme and subsequent evidence~\cite{LIGOScientific:2025rsn}, the intra-model variability of binary black hole waveform approximants is insignificant compared to the inter-model variability of gravitational-wave sources in general. Furthermore, for the binary black hole model, we replicate the analysis performed in~\cite{LIGOScientific:2025rsn} with the only noteworthy difference being calibration error marginalization. This affects the recovered signal-to-noise ratio (SNR) slightly, but has a negligible impact on the parameter estimation; the results presented here for the binary black hole hypothesis are fully consistent with those presented by the LVK in~\cite{LIGOScientific:2025rsn}.

\section{Results}

\begin{figure*}
\includegraphics[width=\columnwidth]{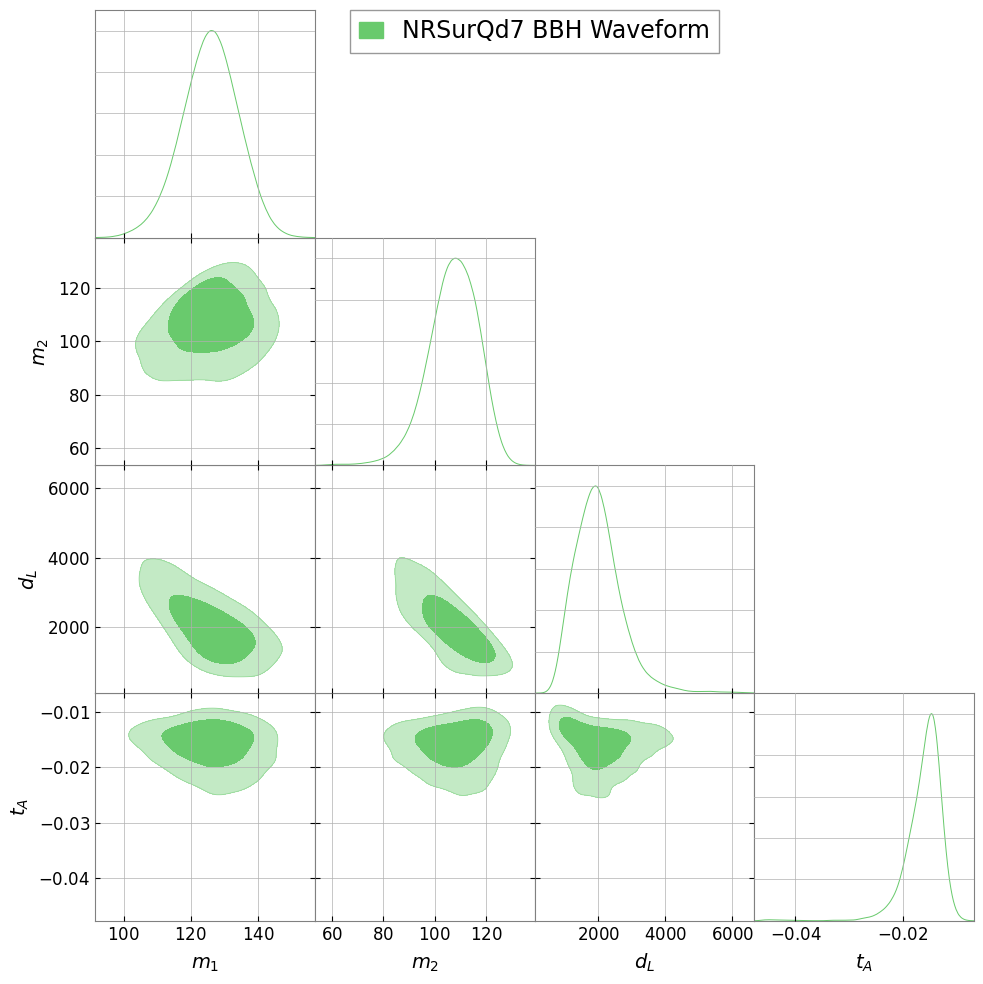}
\caption{A subset of source-frame and extrinsic parameters (component masses, luminosity distance and relative-to-trigger arrival time) under the binary black hole merger event hypothesis with the NRSurQd7 approximant, with 1-$\sigma$ and 2-$\sigma$ contours. }
\label{fig:cornerBBH}
\end{figure*}

Under the binary black hole source hypothesis, using the \texttt{NRSur7Dq4} waveform we recover consistent parameter posteriors with the equivalent waveform analysis of LVK for GW231123~\cite{LIGOScientific:2025rsn}, shown in Fig.~\ref{fig:cornerBBH}. A binary black hole merger with masses of $m_1=125.98^{+13.5}_{-14.6}\,{\rm M_\odot}$ and $m_{2}=107.77^{+14.1}_{-17.3}\,{\rm M_\odot}$, at a luminosity distance $d_L=1932^{+1339}_{-952}\,\text{Mpc}$, provides the highest SNR, as seen in Table~\ref{Tab:BF/SNR}, alongside models' Bayes factors. These binary black hole merger parameter estimates are consistent with the estimates made by the LVK using the \texttt{NRSur7Dq4} waveform alone; $m_1=129^{+15}_{-14}\,{\rm M_\odot}$, and $m_{2}=111^{+14}_{-17}\,{\rm M_\odot}$, $d_L=1500^{+1500}_{-800}\,\text{Mpc}$~\cite{gwoscGW231123}.

\begin{figure*}
\includegraphics[width=0.49\textwidth]{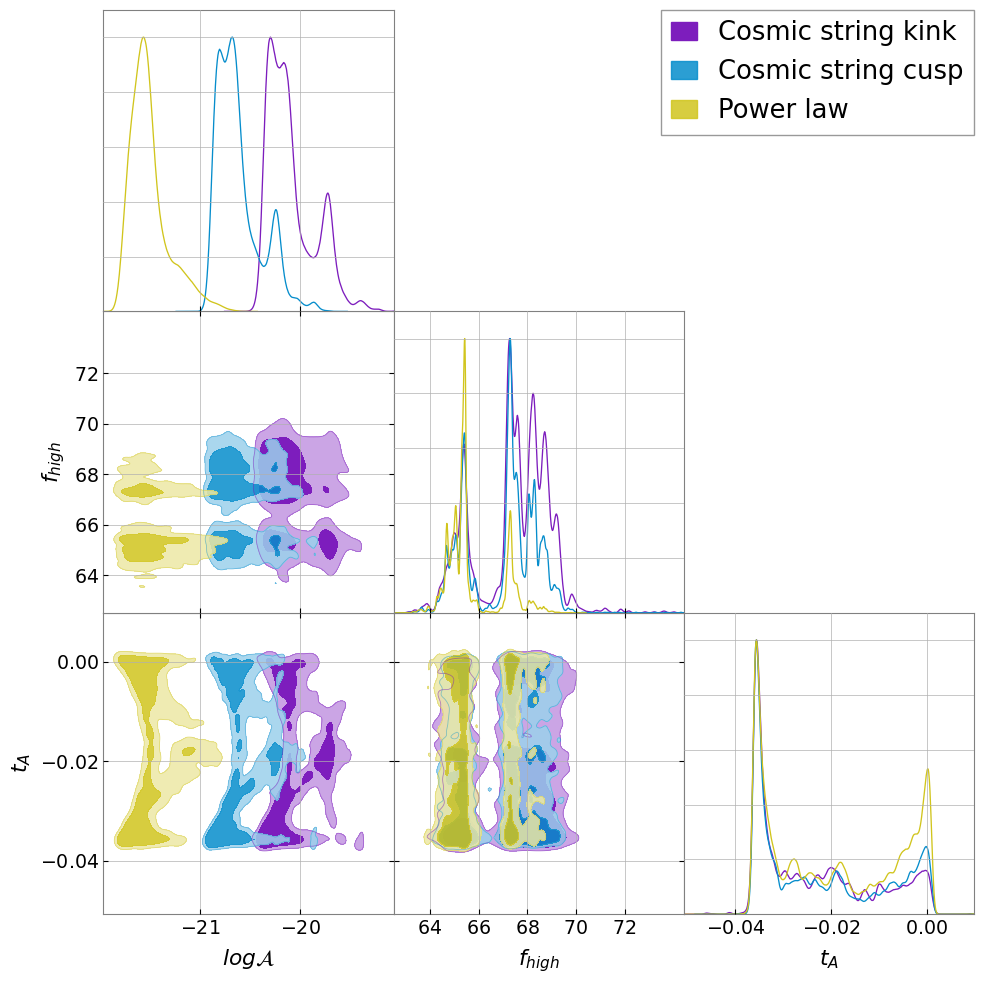}
\includegraphics[width=0.49\textwidth]{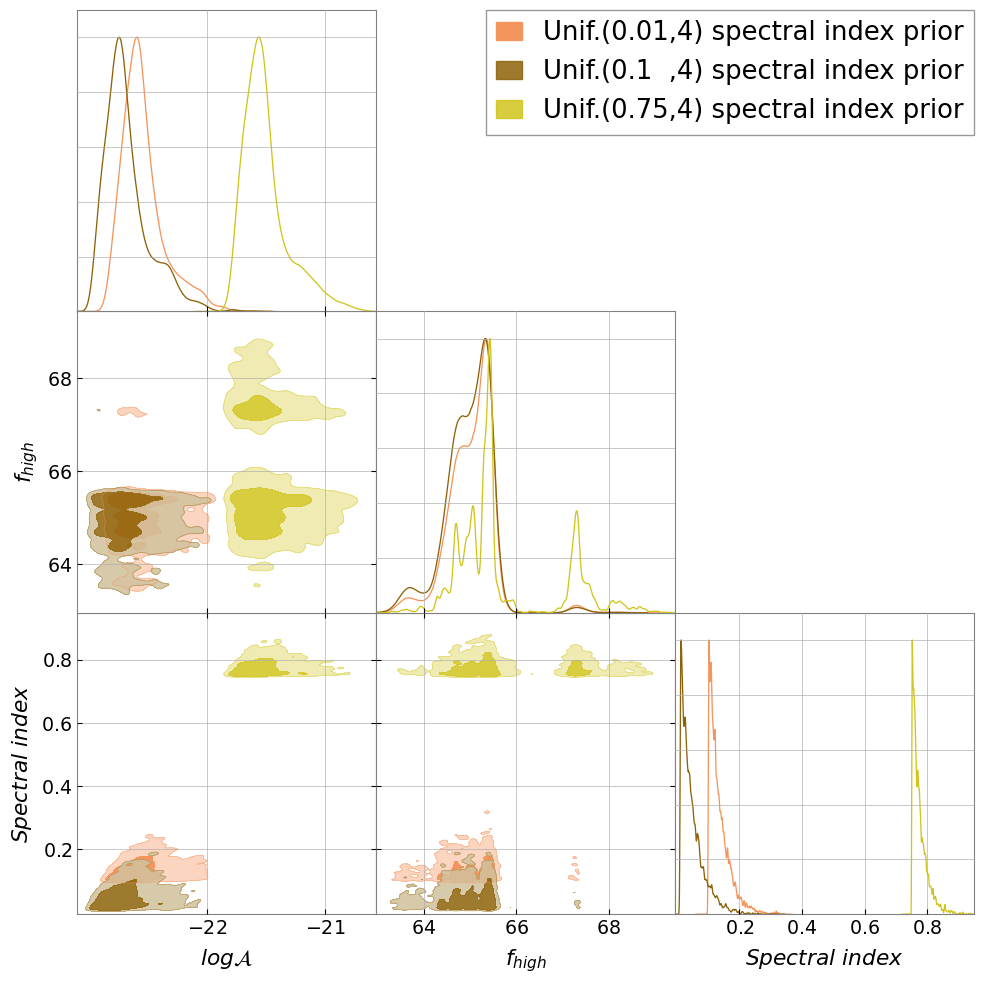}
\caption{Corner plots of posterior samples, for three models considered. Left, cosmic string kink (purple), cosmic string cusp (blue) and arbitrary power law (yellow), the log-amplitude, high frequency cutoff and relative-to-trigger arrival time are shown, with 1-$\sigma$ and 2-$\sigma$ contours. The kink-kink model is also considered and consistent with the above posterior, but not shown here for brevity. Right, the arbitrary power law's spectral index is considered with three different uniform priors (starting at $0.01$ (orange), $0.1$ (brown) and $0.75$ (yellow) respectively), the latter coinciding with the power law on the left. The extra freedom allowed with the power law model shows clear prior-dominated posteriors. }
%\MAB{Q: This is the first time and the only time we are talking about the geocentric time and $t_{trigger}$ appears in the corner plot. In the text only $t_A$ is defined.}
\label{fig:CScorner}
\end{figure*}

For cosmic strings, the cusp, kink, and kink-kink collision cases
result in high frequency cutoff posteriors, $f_{\rm{high},\,\rm{cusp}}=67.3^{+1.6}_{-2.6}$ Hz, $f_{\rm{high},\,\rm{kink}}=67.6^{+1.6}_{-2.8}$ Hz, and $f_{\rm{high},\,\rm{kink-kink}}=67.9^{+3.2}_{-3.2}$ Hz, respectively. They are shown in  Fig.~\ref{fig:CScorner} (left panel), consistent with the merger frequency of a $\sim200\,M_\odot$ binary black hole coalescence. This trend is also observed in the power-law posteriors, displayed in Fig.~\ref{fig:CScorner} (right panel). The power-law model leads to the highest Bayes factor for a source different from a binary black hole merger, compared to noise. This is expected because of the increased degree of freedom, in comparison with the fixed indices in the cusp, kink and kink-kink waveforms. Fig.~\ref{fig:CScorner} (right panel) shows the posteriors for the power-law hypothesis, with different priors for the spectral index. These prior-dominated posteriors, with a clear preference for the lowest spectral index, explain the data preference for the cusp over the kink case.

\begin{table}
\centering
\caption{The $\log_{10}$ Bayes factor between the waveform model considered and the noise (no-signal) hypothesis, as well as network matched filtering signal to noise ratios are reported below. The values for the power law correspond to the $\text{Uniform}\,(0.75,4)$ spectral index prior analysis.}\label{Tab:BF/SNR}
\renewcommand{\arraystretch}{1.2}
\begin{ruledtabular}
\begin{tabular}{lcc}
Waveform Model & $\log\mathcal{BF}$ & SNR\\
\hline
Binary Black Hole & $82.13$ & $21.01^{+0.2}_{-0.3}$\\
Cusp & $38.02$ & $14.66^{+0.1}_{-0.3}$\\
Kink & $30.91$ & $13.59^{+0.1}_{-0.3}$ \\
Kink-kink & $24.40$ &  $12.41^{+0.2}_{-0.4}$\\
Power-law & $45.23$& $16.16^{+0.1}_{-0.3}$\\
%\hline

\end{tabular}
\end{ruledtabular}
\end{table}

\begin{figure*}
\includegraphics[width=\columnwidth]{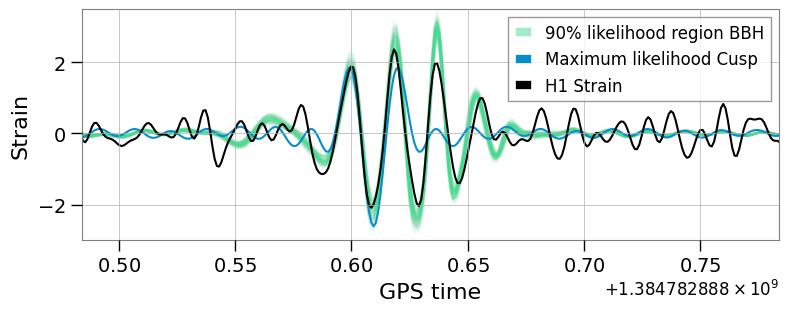}
\caption{Best-fitting cosmic string cusp waveform (blue) and the LIGO Hanford detector strain (black) are shown, whitened with the BayesWave PSD and bandpassed between $20-128\,\text{Hz}$. The $90\%$ confidence region of the NRSur7Dq4 binary black hole waveform template is superimposed (green).}
\label{fig:strains}
\end{figure*}

The mismatch between cosmic strings' prediction 
and data, is shown in Fig.~\ref{fig:strains} in the time-domain. We present the highest likelihood cosmic string cusp match, against the detector strain and 90\% confidence interval for the \texttt{NRSur7Dq4} approximant. The frequency domain power law shape  for cosmic strings, results in a $\sim3$ cycle time domain waveform. These cycles can then be used to fit only around half of the observed strain, which results in a bimodal posterior for the arrival time. This effect is again observed in both cusp and kink cases, to a lesser extent in the latter, as well as in the spectral index-varying generic power law waveform, where it is most prominent. In contrast, the arrival time bimodality is completely non-existent under the binary black hole merger hypothesis, as shown in Fig.~\ref{fig:cornerBBH}.

We note that the cosmic string and power law waveforms considered here do provide an SNR above the $\sim8$ threshold, and decisive evidence against the only-noise hypothesis and therefore the existence of ``a signal''. However, when compared to the best matching binary black hole 
waveform, they are in turn decisively disfavoured. This points out the need to always contextualize source identification with multiple competing hypotheses.

\section{Conclusions}
We have presented a Bayesian model comparison analysis in order to substantiate the source identification of GW231123 as a binary black hole merger~\cite{LIGOScientific:2025rsn}. To this end, we have considered the three main emission channels for gravitational waves produced by cosmic strings in addition to the assumed binary black hole compact binary coalescence event. To further facilitate this comparison, inspired by the power law-like frequency domain representation of all cosmic string waveform models, a freely varying spectral index, generic power law, is also considered. Our analysis is similar to the study of the source of GW190521~\cite{LIGOScientific:2020iuh}, where a Bayesian model compared a binary black hole scenario with a cosmic string origin (cusp and kink), showing preference for the binary black hole~\cite{LIGOScientific:2020ufj}.

For both GW190521 and GW231123 the observed waveform consists of only $\sim 5$ cycles. In the future, events of even shorter duration could be observed. A systematic model comparison test will be important, especially including the numerous possible sources for short duration transient gravitational-wave signals. This is a subject of future work for us.

\begin{acknowledgments}
This research has made use of data~\cite{KAGRA:2023pio} or software obtained from the Gravitational Wave Open Science Center (gwosc.org), a service of the LIGO Scientific Collaboration, the Virgo Collaboration, and KAGRA. This material is based upon work supported by NSF's LIGO Laboratory which is a major facility fully funded by the National Science Foundation, as well as the Science and Technology Facilities Council (STFC) of the United Kingdom, the Max-Planck-Society (MPS), and the State of Niedersachsen/Germany for support of the construction of Advanced LIGO and construction and operation of the GEO600 detector. Additional support for Advanced LIGO was provided by the Australian Research Council. Virgo is funded, through the European Gravitational Observatory (EGO), by the French Centre National de Recherche Scientifique (CNRS), the Italian Istituto Nazionale di Fisica Nucleare (INFN) and the Dutch Nikhef, with contributions by institutions from Belgium, Germany, Greece, Hungary, Ireland, Japan, Monaco, Poland, Portugal, Spain. KAGRA is supported by Ministry of Education, Culture, Sports, Science and Technology (MEXT), Japan Society for the Promotion of Science (JSPS) in Japan; National Research Foundation (NRF) and Ministry of Science and ICT (MSIT) in Korea; Academia Sinica (AS) and National Science and Technology Council (NSTC) in Taiwan.

The authors are grateful for computational resources provided by the LIGO Laboratory and supported by the National Science Foundation Grants PHY-0757058 and PHY-0823459. MS acknowledges support from the Science and Technology Facility Council (STFC), UK, under the research grant ST/X000753/1. This work was supported by the French government through the France 2030 investment plan managed by the National Research Agency (ANR), as part of the Initiative of Excellence of Université Côte d’Azur under reference number ANR-15-IDEX-01. This material is based upon work supported by NSF’s LIGO Laboratory which is a major facility fully funded by the National Science Foundation. This manuscript was assigned LIGO-Document number LIGO-P2500447.

\end{acknowledgments}

\bibliography{apssamp}% Produces the bibliography via BibTeX.

\providecommand{\noopsort}[1]{}\providecommand{\singleletter}[1]{#1}%
\begin{thebibliography}{28}%
\makeatletter
\providecommand \@ifxundefined [1]{%
 \@ifx{#1\undefined}
}%
\providecommand \@ifnum [1]{%
 \ifnum #1\expandafter \@firstoftwo
 \else \expandafter \@secondoftwo
 \fi
}%
\providecommand \@ifx [1]{%
 \ifx #1\expandafter \@firstoftwo
 \else \expandafter \@secondoftwo
 \fi
}%
\providecommand \natexlab [1]{#1}%
\providecommand \enquote  [1]{``#1''}%
\providecommand \bibnamefont  [1]{#1}%
\providecommand \bibfnamefont [1]{#1}%
\providecommand \citenamefont [1]{#1}%
\providecommand \href@noop [0]{\@secondoftwo}%
\providecommand \href [0]{\begingroup \@sanitize@url \@href}%
\providecommand \@href[1]{\@@startlink{#1}\@@href}%
\providecommand \@@href[1]{\endgroup#1\@@endlink}%
\providecommand \@sanitize@url [0]{\catcode `\\12\catcode `\$12\catcode `\&12\catcode `\#12\catcode `\^12\catcode `\_12\catcode `\%12\relax}%
\providecommand \@@startlink[1]{}%
\providecommand \@@endlink[0]{}%
\providecommand \url  [0]{\begingroup\@sanitize@url \@url }%
\providecommand \@url [1]{\endgroup\@href {#1}{\urlprefix }}%
\providecommand \urlprefix  [0]{URL }%
\providecommand \Eprint [0]{\href }%
\providecommand \doibase [0]{https://doi.org/}%
\providecommand \selectlanguage [0]{\@gobble}%
\providecommand \bibinfo  [0]{\@secondoftwo}%
\providecommand \bibfield  [0]{\@secondoftwo}%
\providecommand \translation [1]{[#1]}%
\providecommand \BibitemOpen [0]{}%
\providecommand \bibitemStop [0]{}%
\providecommand \bibitemNoStop [0]{.\EOS\space}%
\providecommand \EOS [0]{\spacefactor3000\relax}%
\providecommand \BibitemShut  [1]{\csname bibitem#1\endcsname}%
\let\auto@bib@innerbib\@empty
%</preamble>
\bibitem [{\citenamefont {Aasi}\ \emph {et~al.}(2015)\citenamefont {Aasi} \emph {et~al.}}]{LIGOScientific:2014pky}%
  \BibitemOpen
  \bibfield  {author} {\bibinfo {author} {\bibfnamefont {J.}~\bibnamefont {Aasi}} \emph {et~al.} (\bibinfo {collaboration} {LIGO Scientific}),\ }\bibfield  {title} {\bibinfo {title} {{Advanced LIGO}},\ }\href {https://doi.org/10.1088/0264-9381/32/7/074001} {\bibfield  {journal} {\bibinfo  {journal} {Class. Quant. Grav.}\ }\textbf {\bibinfo {volume} {32}},\ \bibinfo {pages} {074001} (\bibinfo {year} {2015})},\ \Eprint {https://arxiv.org/abs/1411.4547} {arXiv:1411.4547 [gr-qc]} \BibitemShut {NoStop}%
\bibitem [{\citenamefont {Acernese}\ \emph {et~al.}(2015)\citenamefont {Acernese} \emph {et~al.}}]{VIRGO:2014yos}%
  \BibitemOpen
  \bibfield  {author} {\bibinfo {author} {\bibfnamefont {F.}~\bibnamefont {Acernese}} \emph {et~al.} (\bibinfo {collaboration} {VIRGO}),\ }\bibfield  {title} {\bibinfo {title} {{Advanced Virgo: a second-generation interferometric gravitational wave detector}},\ }\href {https://doi.org/10.1088/0264-9381/32/2/024001} {\bibfield  {journal} {\bibinfo  {journal} {Class. Quant. Grav.}\ }\textbf {\bibinfo {volume} {32}},\ \bibinfo {pages} {024001} (\bibinfo {year} {2015})},\ \Eprint {https://arxiv.org/abs/1408.3978} {arXiv:1408.3978 [gr-qc]} \BibitemShut {NoStop}%
\bibitem [{\citenamefont {Akutsu}\ \emph {et~al.}(2021)\citenamefont {Akutsu} \emph {et~al.}}]{KAGRA:2020tym}%
  \BibitemOpen
  \bibfield  {author} {\bibinfo {author} {\bibfnamefont {T.}~\bibnamefont {Akutsu}} \emph {et~al.} (\bibinfo {collaboration} {KAGRA}),\ }\bibfield  {title} {\bibinfo {title} {{Overview of KAGRA: Detector design and construction history}},\ }\href {https://doi.org/10.1093/ptep/ptaa125} {\bibfield  {journal} {\bibinfo  {journal} {PTEP}\ }\textbf {\bibinfo {volume} {2021}},\ \bibinfo {pages} {05A101} (\bibinfo {year} {2021})},\ \Eprint {https://arxiv.org/abs/2005.05574} {arXiv:2005.05574 [physics.ins-det]} \BibitemShut {NoStop}%
\bibitem [{\citenamefont {LVK}\ \emph {et~al.}(2025)\citenamefont {LVK} \emph {et~al.}}]{LIGOScientific:2025rsn}%
  \BibitemOpen
  \bibfield  {author} {\bibinfo {author} {\bibnamefont {LVK}} \emph {et~al.} (\bibinfo {collaboration} {LIGO Scientific, VIRGO, KAGRA}),\ }\bibfield  {title} {\bibinfo {title} {{GW231123: a Binary Black Hole Merger with Total Mass 190-265 $M_{\odot}$}},\ }\href@noop {} {\bibfield  {journal} {\bibinfo  {journal} {arXiv}\ } (\bibinfo {year} {2025})},\ \Eprint {https://arxiv.org/abs/2507.08219} {arXiv:2507.08219 [astro-ph.HE]} \BibitemShut {NoStop}%
\bibitem [{\citenamefont {Jeannerot}\ \emph {et~al.}(2003)\citenamefont {Jeannerot}, \citenamefont {Rocher},\ and\ \citenamefont {Sakellariadou}}]{Jeannerot:2003qv}%
  \BibitemOpen
  \bibfield  {author} {\bibinfo {author} {\bibfnamefont {R.}~\bibnamefont {Jeannerot}}, \bibinfo {author} {\bibfnamefont {J.}~\bibnamefont {Rocher}},\ and\ \bibinfo {author} {\bibfnamefont {M.}~\bibnamefont {Sakellariadou}},\ }\bibfield  {title} {\bibinfo {title} {{How generic is cosmic string formation in SUSY GUTs}},\ }\href {https://doi.org/10.1103/PhysRevD.68.103514} {\bibfield  {journal} {\bibinfo  {journal} {Phys. Rev. D}\ }\textbf {\bibinfo {volume} {68}},\ \bibinfo {pages} {103514} (\bibinfo {year} {2003})},\ \Eprint {https://arxiv.org/abs/hep-ph/0308134} {arXiv:hep-ph/0308134} \BibitemShut {NoStop}%
\bibitem [{\citenamefont {Abbott}\ \emph {et~al.}(2018)\citenamefont {Abbott} \emph {et~al.}}]{LIGOScientific:2017ikf}%
  \BibitemOpen
  \bibfield  {author} {\bibinfo {author} {\bibfnamefont {B.~P.}\ \bibnamefont {Abbott}} \emph {et~al.} (\bibinfo {collaboration} {LIGO Scientific, Virgo}),\ }\bibfield  {title} {\bibinfo {title} {{Constraints on cosmic strings using data from the first Advanced LIGO observing run}},\ }\href {https://doi.org/10.1103/PhysRevD.97.102002} {\bibfield  {journal} {\bibinfo  {journal} {Phys. Rev. D}\ }\textbf {\bibinfo {volume} {97}},\ \bibinfo {pages} {102002} (\bibinfo {year} {2018})},\ \Eprint {https://arxiv.org/abs/1712.01168} {arXiv:1712.01168 [gr-qc]} \BibitemShut {NoStop}%
\bibitem [{\citenamefont {Abbott}\ \emph {et~al.}(2021)\citenamefont {Abbott} \emph {et~al.}}]{LIGOScientific:2021nrg}%
  \BibitemOpen
  \bibfield  {author} {\bibinfo {author} {\bibfnamefont {R.}~\bibnamefont {Abbott}} \emph {et~al.} (\bibinfo {collaboration} {LIGO Scientific, Virgo, KAGRA}),\ }\bibfield  {title} {\bibinfo {title} {{Constraints on Cosmic Strings Using Data from the Third Advanced LIGO{\textendash}Virgo Observing Run}},\ }\href {https://doi.org/10.1103/PhysRevLett.126.241102} {\bibfield  {journal} {\bibinfo  {journal} {Phys. Rev. Lett.}\ }\textbf {\bibinfo {volume} {126}},\ \bibinfo {pages} {241102} (\bibinfo {year} {2021})},\ \Eprint {https://arxiv.org/abs/2101.12248} {arXiv:2101.12248 [gr-qc]} \BibitemShut {NoStop}%
\bibitem [{\citenamefont {Damour}\ and\ \citenamefont {Vilenkin}(2000)}]{Damour:2000wa}%
  \BibitemOpen
  \bibfield  {author} {\bibinfo {author} {\bibfnamefont {T.}~\bibnamefont {Damour}}\ and\ \bibinfo {author} {\bibfnamefont {A.}~\bibnamefont {Vilenkin}},\ }\bibfield  {title} {\bibinfo {title} {{Gravitational wave bursts from cosmic strings}},\ }\href {https://doi.org/10.1103/PhysRevLett.85.3761} {\bibfield  {journal} {\bibinfo  {journal} {Phys. Rev. Lett.}\ }\textbf {\bibinfo {volume} {85}},\ \bibinfo {pages} {3761} (\bibinfo {year} {2000})},\ \Eprint {https://arxiv.org/abs/gr-qc/0004075} {arXiv:gr-qc/0004075} \BibitemShut {NoStop}%
\bibitem [{\citenamefont {Damour}\ and\ \citenamefont {Vilenkin}(2001)}]{Damour:2001bk}%
  \BibitemOpen
  \bibfield  {author} {\bibinfo {author} {\bibfnamefont {T.}~\bibnamefont {Damour}}\ and\ \bibinfo {author} {\bibfnamefont {A.}~\bibnamefont {Vilenkin}},\ }\bibfield  {title} {\bibinfo {title} {{Gravitational wave bursts from cusps and kinks on cosmic strings}},\ }\href {https://doi.org/10.1103/PhysRevD.64.064008} {\bibfield  {journal} {\bibinfo  {journal} {Phys. Rev. D}\ }\textbf {\bibinfo {volume} {64}},\ \bibinfo {pages} {064008} (\bibinfo {year} {2001})},\ \Eprint {https://arxiv.org/abs/gr-qc/0104026} {arXiv:gr-qc/0104026} \BibitemShut {NoStop}%
\bibitem [{\citenamefont {Damour}\ and\ \citenamefont {Vilenkin}(2005)}]{Damour:2004kw}%
  \BibitemOpen
  \bibfield  {author} {\bibinfo {author} {\bibfnamefont {T.}~\bibnamefont {Damour}}\ and\ \bibinfo {author} {\bibfnamefont {A.}~\bibnamefont {Vilenkin}},\ }\bibfield  {title} {\bibinfo {title} {{Gravitational radiation from cosmic (super)strings: Bursts, stochastic background, and observational windows}},\ }\href {https://doi.org/10.1103/PhysRevD.71.063510} {\bibfield  {journal} {\bibinfo  {journal} {Phys. Rev. D}\ }\textbf {\bibinfo {volume} {71}},\ \bibinfo {pages} {063510} (\bibinfo {year} {2005})},\ \Eprint {https://arxiv.org/abs/hep-th/0410222} {arXiv:hep-th/0410222} \BibitemShut {NoStop}%
\bibitem [{\citenamefont {Bennett}\ and\ \citenamefont {Bouchet}(1988)}]{PhysRevLett.60.257}%
  \BibitemOpen
  \bibfield  {author} {\bibinfo {author} {\bibfnamefont {D.~P.}\ \bibnamefont {Bennett}}\ and\ \bibinfo {author} {\bibfnamefont {F.~m. c.~R.}\ \bibnamefont {Bouchet}},\ }\bibfield  {title} {\bibinfo {title} {Evidence for a scaling solution in cosmic-string evolution},\ }\href {https://doi.org/10.1103/PhysRevLett.60.257} {\bibfield  {journal} {\bibinfo  {journal} {Phys. Rev. Lett.}\ }\textbf {\bibinfo {volume} {60}},\ \bibinfo {pages} {257} (\bibinfo {year} {1988})}\BibitemShut {NoStop}%
\bibitem [{\citenamefont {Allen}\ and\ \citenamefont {Shellard}(1990)}]{PhysRevLett.64.119}%
  \BibitemOpen
  \bibfield  {author} {\bibinfo {author} {\bibfnamefont {B.}~\bibnamefont {Allen}}\ and\ \bibinfo {author} {\bibfnamefont {E.~P.~S.}\ \bibnamefont {Shellard}},\ }\bibfield  {title} {\bibinfo {title} {Cosmic-string evolution: A numerical simulation},\ }\href {https://doi.org/10.1103/PhysRevLett.64.119} {\bibfield  {journal} {\bibinfo  {journal} {Phys. Rev. Lett.}\ }\textbf {\bibinfo {volume} {64}},\ \bibinfo {pages} {119} (\bibinfo {year} {1990})}\BibitemShut {NoStop}%
\bibitem [{\citenamefont {Sakellariadou}\ and\ \citenamefont {Vilenkin}(1990)}]{PhysRevD.42.349}%
  \BibitemOpen
  \bibfield  {author} {\bibinfo {author} {\bibfnamefont {M.}~\bibnamefont {Sakellariadou}}\ and\ \bibinfo {author} {\bibfnamefont {A.}~\bibnamefont {Vilenkin}},\ }\bibfield  {title} {\bibinfo {title} {Cosmic-string evolution in flat spacetime},\ }\href {https://doi.org/10.1103/PhysRevD.42.349} {\bibfield  {journal} {\bibinfo  {journal} {Phys. Rev. D}\ }\textbf {\bibinfo {volume} {42}},\ \bibinfo {pages} {349} (\bibinfo {year} {1990})}\BibitemShut {NoStop}%
\bibitem [{\citenamefont {Sakellariadou}(1990)}]{Sakellariadou:1990ne}%
  \BibitemOpen
  \bibfield  {author} {\bibinfo {author} {\bibfnamefont {M.}~\bibnamefont {Sakellariadou}},\ }\bibfield  {title} {\bibinfo {title} {{Gravitational waves emitted from infinite strings}},\ }\href {https://doi.org/10.1103/PhysRevD.42.354} {\bibfield  {journal} {\bibinfo  {journal} {Phys. Rev. D}\ }\textbf {\bibinfo {volume} {42}},\ \bibinfo {pages} {354} (\bibinfo {year} {1990})},\ \bibinfo {note} {[Erratum: Phys.Rev.D 43, 4150 (1991)]}\BibitemShut {NoStop}%
\bibitem [{\citenamefont {Abbott}\ \emph {et~al.}(2020{\natexlab{a}})\citenamefont {Abbott} \emph {et~al.}}]{LIGOScientific:2020iuh}%
  \BibitemOpen
  \bibfield  {author} {\bibinfo {author} {\bibfnamefont {R.}~\bibnamefont {Abbott}} \emph {et~al.} (\bibinfo {collaboration} {LIGO Scientific, Virgo}),\ }\bibfield  {title} {\bibinfo {title} {{GW190521: A Binary Black Hole Merger with a Total Mass of $150 M_{\odot}$}},\ }\href {https://doi.org/10.1103/PhysRevLett.125.101102} {\bibfield  {journal} {\bibinfo  {journal} {Phys. Rev. Lett.}\ }\textbf {\bibinfo {volume} {125}},\ \bibinfo {pages} {101102} (\bibinfo {year} {2020}{\natexlab{a}})},\ \Eprint {https://arxiv.org/abs/2009.01075} {arXiv:2009.01075 [gr-qc]} \BibitemShut {NoStop}%
\bibitem [{\citenamefont {Abbott}\ \emph {et~al.}(2020{\natexlab{b}})\citenamefont {Abbott} \emph {et~al.}}]{LIGOScientific:2020ufj}%
  \BibitemOpen
  \bibfield  {author} {\bibinfo {author} {\bibfnamefont {R.}~\bibnamefont {Abbott}} \emph {et~al.} (\bibinfo {collaboration} {LIGO Scientific, Virgo}),\ }\bibfield  {title} {\bibinfo {title} {{Properties and Astrophysical Implications of the 150 M$_\odot$ Binary Black Hole Merger GW190521}},\ }\href {https://doi.org/10.3847/2041-8213/aba493} {\bibfield  {journal} {\bibinfo  {journal} {Astrophys. J. Lett.}\ }\textbf {\bibinfo {volume} {900}},\ \bibinfo {pages} {L13} (\bibinfo {year} {2020}{\natexlab{b}})},\ \Eprint {https://arxiv.org/abs/2009.01190} {arXiv:2009.01190 [astro-ph.HE]} \BibitemShut {NoStop}%
\bibitem [{\citenamefont {Aurrekoetxea}\ \emph {et~al.}(2024)\citenamefont {Aurrekoetxea}, \citenamefont {Hoy},\ and\ \citenamefont {Hannam}}]{Aurrekoetxea:2023vtp}%
  \BibitemOpen
  \bibfield  {author} {\bibinfo {author} {\bibfnamefont {J.~C.}\ \bibnamefont {Aurrekoetxea}}, \bibinfo {author} {\bibfnamefont {C.}~\bibnamefont {Hoy}},\ and\ \bibinfo {author} {\bibfnamefont {M.}~\bibnamefont {Hannam}},\ }\bibfield  {title} {\bibinfo {title} {{Revisiting the Cosmic String Origin of GW190521}},\ }\href {https://doi.org/10.1103/PhysRevLett.132.181401} {\bibfield  {journal} {\bibinfo  {journal} {Phys. Rev. Lett.}\ }\textbf {\bibinfo {volume} {132}},\ \bibinfo {pages} {181401} (\bibinfo {year} {2024})},\ \Eprint {https://arxiv.org/abs/2312.03860} {arXiv:2312.03860 [gr-qc]} \BibitemShut {NoStop}%
\bibitem [{\citenamefont {LVK}()}]{gwoscGW231123}%
  \BibitemOpen
  \bibfield  {author} {\bibinfo {author} {\bibnamefont {LVK}},\ }\href {https://doi.org/10.7935/anj7-6q40} {\bibinfo {title} {{GWOSC DATA GW231123, doi.org/10.7935/anj7-6q40}}}\BibitemShut {NoStop}%
\bibitem [{\citenamefont {Abbott}\ \emph {et~al.}(2023)\citenamefont {Abbott} \emph {et~al.}}]{KAGRA:2023pio}%
  \BibitemOpen
  \bibfield  {author} {\bibinfo {author} {\bibfnamefont {R.}~\bibnamefont {Abbott}} \emph {et~al.} (\bibinfo {collaboration} {KAGRA, VIRGO, LIGO Scientific}),\ }\bibfield  {title} {\bibinfo {title} {{Open Data from the Third Observing Run of LIGO, Virgo, KAGRA, and GEO}},\ }\href {https://doi.org/10.3847/1538-4365/acdc9f} {\bibfield  {journal} {\bibinfo  {journal} {Astrophys. J. Suppl.}\ }\textbf {\bibinfo {volume} {267}},\ \bibinfo {pages} {29} (\bibinfo {year} {2023})},\ \Eprint {https://arxiv.org/abs/2302.03676} {arXiv:2302.03676 [gr-qc]} \BibitemShut {NoStop}%
\bibitem [{\citenamefont {Christensen}\ and\ \citenamefont {Meyer}(2022)}]{Christensen:2022bxb}%
  \BibitemOpen
  \bibfield  {author} {\bibinfo {author} {\bibfnamefont {N.}~\bibnamefont {Christensen}}\ and\ \bibinfo {author} {\bibfnamefont {R.}~\bibnamefont {Meyer}},\ }\bibfield  {title} {\bibinfo {title} {{Parameter estimation with gravitational waves}},\ }\href {https://doi.org/10.1103/RevModPhys.94.025001} {\bibfield  {journal} {\bibinfo  {journal} {Rev. Mod. Phys.}\ }\textbf {\bibinfo {volume} {94}},\ \bibinfo {pages} {025001} (\bibinfo {year} {2022})},\ \Eprint {https://arxiv.org/abs/2204.04449} {arXiv:2204.04449 [gr-qc]} \BibitemShut {NoStop}%
\bibitem [{\citenamefont {{Skilling}}(2004)}]{2004AIPC..735..395S}%
  \BibitemOpen
  \bibfield  {author} {\bibinfo {author} {\bibfnamefont {J.}~\bibnamefont {{Skilling}}},\ }\bibfield  {title} {\bibinfo {title} {{Nested Sampling}},\ }in\ \href {https://doi.org/10.1063/1.1835238} {\emph {\bibinfo {booktitle} {Bayesian Inference and Maximum Entropy Methods in Science and Engineering: 24th International Workshop on Bayesian Inference and Maximum Entropy Methods in Science and Engineering}}},\ \bibinfo {series} {American Institute of Physics Conference Series}, Vol.\ \bibinfo {volume} {735},\ \bibinfo {editor} {edited by\ \bibinfo {editor} {\bibfnamefont {R.}~\bibnamefont {{Fischer}}}, \bibinfo {editor} {\bibfnamefont {R.}~\bibnamefont {{Preuss}}},\ and\ \bibinfo {editor} {\bibfnamefont {U.~V.}\ \bibnamefont {{Toussaint}}}}\ (\bibinfo {year} {2004})\ pp.\ \bibinfo {pages} {395--405}\BibitemShut {NoStop}%
\bibitem [{\citenamefont {Skilling}(2006)}]{10.1214/06-BA127}%
  \BibitemOpen
  \bibfield  {author} {\bibinfo {author} {\bibfnamefont {J.}~\bibnamefont {Skilling}},\ }\bibfield  {title} {\bibinfo {title} {{Nested sampling for general Bayesian computation}},\ }\href {https://doi.org/10.1214/06-BA127} {\bibfield  {journal} {\bibinfo  {journal} {Bayesian Analysis}\ }\textbf {\bibinfo {volume} {1}},\ \bibinfo {pages} {833 } (\bibinfo {year} {2006})}\BibitemShut {NoStop}%
\bibitem [{\citenamefont {{Speagle}}(2020)}]{2020MNRAS.493.3132S}%
  \BibitemOpen
  \bibfield  {author} {\bibinfo {author} {\bibfnamefont {J.~S.}\ \bibnamefont {{Speagle}}},\ }\bibfield  {title} {\bibinfo {title} {{DYNESTY: a dynamic nested sampling package for estimating Bayesian posteriors and evidences}},\ }\href {https://doi.org/10.1093/mnras/staa278} {\bibfield  {journal} {\bibinfo  {journal} {Mon. Not. Roy. Astron. Soc.}\ }\textbf {\bibinfo {volume} {493}},\ \bibinfo {pages} {3132} (\bibinfo {year} {2020})},\ \Eprint {https://arxiv.org/abs/1904.02180} {arXiv:1904.02180 [astro-ph.IM]} \BibitemShut {NoStop}%
\bibitem [{\citenamefont {Koposov}\ \emph {et~al.}(2023)\citenamefont {Koposov}, \citenamefont {Speagle}, \citenamefont {Barbary}, \citenamefont {Ashton}, \citenamefont {Bennett}, \citenamefont {Buchner}, \citenamefont {Scheffler}, \citenamefont {Cook}, \citenamefont {Talbot}, \citenamefont {Guillochon}, \citenamefont {Cubillos}, \citenamefont {Ramos}, \citenamefont {Johnson}, \citenamefont {Lang}, \citenamefont {Ilya}, \citenamefont {Dartiailh}, \citenamefont {Nitz}, \citenamefont {McCluskey}, \citenamefont {Archibald}, \citenamefont {Deil}, \citenamefont {Foreman-Mackey}, \citenamefont {Goldstein}, \citenamefont {Tollerud}, \citenamefont {Leja}, \citenamefont {Kirk}, \citenamefont {Pitkin}, \citenamefont {Sheehan}, \citenamefont {Cargile}, \citenamefont {Patel},\ and\ \citenamefont {Angus}}]{sergey_koposov_2023_7600689}%
  \BibitemOpen
  \bibfield  {author} {\bibinfo {author} {\bibfnamefont {S.}~\bibnamefont {Koposov}}, \bibinfo {author} {\bibfnamefont {J.}~\bibnamefont {Speagle}}, \bibinfo {author} {\bibfnamefont {K.}~\bibnamefont {Barbary}}, \bibinfo {author} {\bibfnamefont {G.}~\bibnamefont {Ashton}}, \bibinfo {author} {\bibfnamefont {E.}~\bibnamefont {Bennett}}, \bibinfo {author} {\bibfnamefont {J.}~\bibnamefont {Buchner}}, \bibinfo {author} {\bibfnamefont {C.}~\bibnamefont {Scheffler}}, \bibinfo {author} {\bibfnamefont {B.}~\bibnamefont {Cook}}, \bibinfo {author} {\bibfnamefont {C.}~\bibnamefont {Talbot}}, \bibinfo {author} {\bibfnamefont {J.}~\bibnamefont {Guillochon}}, \bibinfo {author} {\bibfnamefont {P.}~\bibnamefont {Cubillos}}, \bibinfo {author} {\bibfnamefont {A.~A.}\ \bibnamefont {Ramos}}, \bibinfo {author} {\bibfnamefont {B.}~\bibnamefont {Johnson}}, \bibinfo {author} {\bibfnamefont {D.}~\bibnamefont {Lang}}, \bibinfo {author} {\bibnamefont {Ilya}}, \bibinfo {author} {\bibfnamefont {M.}~\bibnamefont {Dartiailh}}, \bibinfo
  {author} {\bibfnamefont {A.}~\bibnamefont {Nitz}}, \bibinfo {author} {\bibfnamefont {A.}~\bibnamefont {McCluskey}}, \bibinfo {author} {\bibfnamefont {A.}~\bibnamefont {Archibald}}, \bibinfo {author} {\bibfnamefont {C.}~\bibnamefont {Deil}}, \bibinfo {author} {\bibfnamefont {D.}~\bibnamefont {Foreman-Mackey}}, \bibinfo {author} {\bibfnamefont {D.}~\bibnamefont {Goldstein}}, \bibinfo {author} {\bibfnamefont {E.}~\bibnamefont {Tollerud}}, \bibinfo {author} {\bibfnamefont {J.}~\bibnamefont {Leja}}, \bibinfo {author} {\bibfnamefont {M.}~\bibnamefont {Kirk}}, \bibinfo {author} {\bibfnamefont {M.}~\bibnamefont {Pitkin}}, \bibinfo {author} {\bibfnamefont {P.}~\bibnamefont {Sheehan}}, \bibinfo {author} {\bibfnamefont {P.}~\bibnamefont {Cargile}}, \bibinfo {author} {\bibfnamefont {R.}~\bibnamefont {Patel}},\ and\ \bibinfo {author} {\bibfnamefont {R.}~\bibnamefont {Angus}},\ }\href {https://doi.org/10.5281/zenodo.7600689} {\bibinfo {title} {joshspeagle/dynesty: v2.1.0}} (\bibinfo {year} {2023})\BibitemShut {NoStop}%
\bibitem [{\citenamefont {Ashton}\ \emph {et~al.}(2019)\citenamefont {Ashton} \emph {et~al.}}]{2019ApJS..241...27A}%
  \BibitemOpen
  \bibfield  {author} {\bibinfo {author} {\bibfnamefont {G.}~\bibnamefont {Ashton}} \emph {et~al.},\ }\bibfield  {title} {\bibinfo {title} {{BILBY: A user-friendly Bayesian inference library for gravitational-wave astronomy}},\ }\href {https://doi.org/10.3847/1538-4365/ab06fc} {\bibfield  {journal} {\bibinfo  {journal} {Astrophys. J. Suppl.}\ }\textbf {\bibinfo {volume} {241}},\ \bibinfo {pages} {27} (\bibinfo {year} {2019})},\ \Eprint {https://arxiv.org/abs/1811.02042} {arXiv:1811.02042 [astro-ph.IM]} \BibitemShut {NoStop}%
\bibitem [{\citenamefont {Romero-Shaw}\ \emph {et~al.}(2020)\citenamefont {Romero-Shaw} \emph {et~al.}}]{2020MNRAS.499.3295R}%
  \BibitemOpen
  \bibfield  {author} {\bibinfo {author} {\bibfnamefont {I.~M.}\ \bibnamefont {Romero-Shaw}} \emph {et~al.},\ }\bibfield  {title} {\bibinfo {title} {{Bayesian inference for compact binary coalescences with bilby: validation and application to the first LIGO{\textendash}Virgo gravitational-wave transient catalogue}},\ }\href {https://doi.org/10.1093/mnras/staa2850} {\bibfield  {journal} {\bibinfo  {journal} {Mon. Not. Roy. Astron. Soc.}\ }\textbf {\bibinfo {volume} {499}},\ \bibinfo {pages} {3295} (\bibinfo {year} {2020})},\ \Eprint {https://arxiv.org/abs/2006.00714} {arXiv:2006.00714 [astro-ph.IM]} \BibitemShut {NoStop}%
\bibitem [{\citenamefont {Divakarla}\ \emph {et~al.}(2020)\citenamefont {Divakarla}, \citenamefont {Thrane}, \citenamefont {Lasky},\ and\ \citenamefont {Whiting}}]{Divakarla:2019zjj}%
  \BibitemOpen
  \bibfield  {author} {\bibinfo {author} {\bibfnamefont {A.~K.}\ \bibnamefont {Divakarla}}, \bibinfo {author} {\bibfnamefont {E.}~\bibnamefont {Thrane}}, \bibinfo {author} {\bibfnamefont {P.~D.}\ \bibnamefont {Lasky}},\ and\ \bibinfo {author} {\bibfnamefont {B.~F.}\ \bibnamefont {Whiting}},\ }\bibfield  {title} {\bibinfo {title} {{Memory Effect or Cosmic String? Classifying Gravitational-Wave Bursts with Bayesian Inference}},\ }\href {https://doi.org/10.1103/PhysRevD.102.023010} {\bibfield  {journal} {\bibinfo  {journal} {Phys. Rev. D}\ }\textbf {\bibinfo {volume} {102}},\ \bibinfo {pages} {023010} (\bibinfo {year} {2020})},\ \Eprint {https://arxiv.org/abs/1911.07998} {arXiv:1911.07998 [gr-qc]} \BibitemShut {NoStop}%
\bibitem [{\citenamefont {Varma}\ \emph {et~al.}(2019)\citenamefont {Varma} \emph {et~al.}}]{Varma:2019csw}%
  \BibitemOpen
  \bibfield  {author} {\bibinfo {author} {\bibfnamefont {V.}~\bibnamefont {Varma}} \emph {et~al.},\ }\bibfield  {title} {\bibinfo {title} {{Surrogate models for precessing binary black hole simulations with unequal masses}},\ }\href {https://doi.org/10.1103/PhysRevResearch.1.033015} {\bibfield  {journal} {\bibinfo  {journal} {Phys. Rev. Research.}\ }\textbf {\bibinfo {volume} {1}},\ \bibinfo {pages} {033015} (\bibinfo {year} {2019})},\ \Eprint {https://arxiv.org/abs/1905.09300} {arXiv:1905.09300 [gr-qc]} \BibitemShut {NoStop}%
\end{thebibliography}%

\end{document}